\documentclass[11pt]{article}
\usepackage{moriond,epsfig}

\def\La{\Lambda}
\def\Al{\overline\Lambda}
\def\XI{\Xi^{-}}

\def\be{\begin{equation}}
\def\ee{\end{equation}}
\def\bea{\begin{eqnarray}}
\def\eea{\end{eqnarray}}

\begin{document}
\vspace*{4cm}
\title{New results from the NA57 experiment}

\author{Presented by G.E. Bruno for the NA57 Collaboration:
 \\
\vspace{2mm}
F.~Antinori$^{l}$,
P.A.~Bacon$^{e}$,
A.~Badal{\`a}$^{g}$,
R.~Barbera$^{g}$,
A.~Belogianni$^{a}$,
A.~Bhasin$^{e}$,
I.J.~Bloodworth$^{e}$,
M.~Bombara$^{j}$,
G.E.~Bruno$^{b}$,
S.A.~Bull$^{e}$,
R.~Caliandro$^{b}$,
M.~Campbell$^{h}$,
W.~Carena$^{h}$,
N.~Carrer$^{h}$,
R.F.~Clarke$^{e}$,
A.~Dainese$^{l}$,
A.P.~de~Haas$^{s}$,
P.C.~de~Rijke$^{s}$,
D.~Di~Bari$^{b}$,
S.~Di~Liberto$^{o}$,
R.~Divia$^{h}$,
D.~Elia$^{b}$,
D.~Evans$^{e}$,
K.~Fanebust$^{c}$,
F.~Fayazzadeh$^{k}$,
G.A.~Feofilov$^{q}$,
R.A.~Fini$^{b}$,
P. Ganoti$^{a}$,
B.~Ghidini$^{b}$,
G.~Grella$^{p}$,
H.~Helstrup$^{d}$,
M.~Henriquez$^{k}$,
A.K.~Holme$^{k}$,
A.~Jacholkowski$^{b}$,
G.T.~Jones$^{e}$,
P.~Jovanovic$^{e}$,
A.~Jusko$^{i}$,
R.~Kamermans$^{s}$,
J.B.~Kinson$^{e}$,
K.~Knudson$^{h}$,
A.A.~Kolozhvari$^{q}$,
V.~Kondratiev$^{q}$,
I.~Kr\'alik$^{i}$,
A.~Krav\v c\'akov\'a$^{j}$,
P.~Kuijer$^{s}$,
V.~Lenti$^{b}$,
R.~Lietava$^{f}$,
G.~L\o vh\o iden$^{k}$,
V.~Manzari$^{b}$,
G.~Martinsk\'a$^{j}$,
M.A.~Mazzoni$^{o}$,
F.~Meddi$^{o}$,
A.~Michalon$^{r}$,
M.~Morando$^{l}$,
E.~Nappi$^{b}$,
F.~Navach$^{b}$,
P.I.~Norman$^{e}$,
A.~Palmeri$^{g}$,
G.S.~Pappalardo$^{g}$,
B.~Pastir\v c\'ak$^{i}$,
J.~Pi\v s\'ut$^{f}$,
N.~Pisutova$^{f}$,
F.~Posa$^{b}$,
E.~Quercigh$^{l}$,
F.~Riggi$^{g}$,
D.~R\"ohrich$^{c}$,
G.~Romano$^{p}$,
K.~\v{S}afa\v{r}\'{i}k$^{h}$,
L.~\v S\'andor$^{h,i}$,
E.~Schillings$^{s}$,
G.~Segato$^{l}$,
M.~Sen\`e$^{m}$,
R.~Sen\`e$^{m}$,
W.~Snoeys$^{h}$,
F.~Soramel$^{l}$,
M.~Spyropoulou-Stassinaki$^{a}$,
P.~Staroba$^{n}$,
T.A.~Toulina$^{q}$,
R.~Turrisi$^{l}$,
T.S.~Tveter$^{k}$,
J.~Urb\'{a}n$^{j}$,
F.F.~Valiev$^{q}$,
A.~van~den~Brink$^{s}$,
P.~van~de~Ven$^{s}$,
P. Vande Vyvre$^{h}$,
N.~van~Eijndhoven$^{s}$,
J.~van~Hunen$^{h}$,
A.~Vascotto$^{h}$,
T.~Vik$^{k}$,
O.~Villalobos Baillie$^{e}$,
L.~Vinogradov$^{q}$,
T.~Virgili$^{p}$,
M.F.~Votruba$^{e}$,
J.~Vrl\'{a}kov\'{a}$^{j}$ and
P.~Z\'{a}vada$^{n}$
\vspace{2mm}
}
\address{$^{a}$ Physics Department, University of Athens, Athens, Greece\\
$^{b}$ Dipartimento IA di Fisica dell'Universit{\`a}
  e del Politecnico di Bari and INFN, Bari, Italy \\
$^{c}$ Fysisk Institutt, Universitetet i Bergen, Bergen, Norway\\
$^{d}$ H{\o}gskolen i Bergen, Bergen, Norway\\
$^{e}$ University of Birmingham, Birmingham, UK\\
$^{f}$ Comenius University, Bratislava, Slovakia\\
$^{g}$ University of Catania and INFN, Catania, Italy\\
$^{h}$ CERN, European Laboratory for Particle Physics, Geneva,
       Switzerland\\
$^{i}$ Institute of Experimental Physics, Slovak Academy of Science,
       Ko\v{s}ice, Slovakia\\
$^{j}$ P.J. \v{S}af\'{a}rik University, Ko\v{s}ice, Slovakia\\
$^{k}$ Fysisk Institutt, Universitetet i Oslo, Oslo, Norway\\
$^{l}$ University of Padua and INFN, Padua, Italy\\
$^{m}$ Coll\`ege de France, Paris, France\\
$^{n}$ Institute of Physics, Prague, Czech Republic\\
$^{o}$ University ``La Sapienza'' and INFN, Rome, Italy\\
$^{p}$ Dipartimento di Scienze Fisiche ``E.R. Caianiello''
       dell'Universit{\`a} and INFN, Salerno, Italy\\
$^{q}$ State University of St. Petersburg, St. Petersburg, Russia\\
$^{r}$ Institut de Recherches Subatomique, IN2P3/ULP, Strasbourg, France\\
$^{s}$ Utrecht University and NIKHEF, Utrecht, The Netherlands
}

\maketitle

\abstracts{
We report results from the experiment NA57 at CERN SPS on hyperon production 
at midrapidity in Pb-Pb collisions at 158 $A$\ GeV/$c$\ and 40 $A$\ GeV/$c$. 
$\Lambda$, $\Xi$\ and $\Omega$\ yields are compared with those from the STAR 
experiment at the higher energy of the BNL RHIC. 
$\Lambda$, $\Xi$,  $\Omega$\ and preliminary $K_S^0$\ transverse mass 
spectra are presented and interpreted within the framework of a  
hydro-dynamical blast wave model.}


\setcounter{figure}{0}

\section{Introduction}
The NA57 experiment has been designed to study the onset of enhanced
production of multi-strange baryons and anti-baryons
in Pb-Pb collisions with respect to p-Be collisions. This enhancement, 
first observed by experiment WA97~\cite{Ant99}, is considered as 
evidence~\cite{RafMul82} for a phase transition to a new 
state of matter -- the 
Quark Gluon Plasma (QGP).  \newline
The assessment of the combined results of the CERN heavy-ion experiments 
suggests indeed that a colour deconfined new state of 
matter~\cite{PresReal} is produced in  
central Pb-Pb collisions at 158 $A$~GeV/$c$. 

\noindent
NA57 has extend the study initiated by its predecessor WA97 
by investigating the dependence of the hyperon production  
{\em (i)} on the interaction volume and {\em (ii)}  on the collision 
energy per incoming nucleon~\cite{NA57prop}. 
For the first purpose, special efforts were made in NA57 to enlarge, with 
respect to WA97, the triggered fraction of the total inelastic cross-section 
thus extending the centrality range towards less central collisions;  
for the second,  the experiment has collected data at two beam momenta: 
158 $A$\ GeV/$c$\ and 40 $A$\ GeV/$c$.  

\section{The NA57 experiment}
The NA57 apparatus has been described in detail elsewhere~\cite{Virgili01}. 
Strange and multi-strange hyperons are identified by reconstructing their weak 
decays into final states containing only charged particles, 
e.g.: $\Xi^-$ $\rightarrow$ $\Lambda\pi^-$, with 
$\Lambda$ $\rightarrow$ ${p}\pi^-$.  
Tracks are measured in the silicon 
telescope, an array of pixel detector planes with $5 \times 5$\ ${\rm cm}^2$\ 
cross section, having a total length of about 30 cm. To improve the momentum 
resolution of high momentum tracks an array of double-sided silicon microstrip 
detectors is placed downstream of the tracking telescope. 
The whole silicon telescope is placed inside 
a 1.4 Tesla magnetic field,    
above the beam line, inclined and aligned 
with the lower edge of the detectors laying on a line pointing back to the target. 
The inclination angle $\alpha$\ with respect to the beam line 
and the distance $d$\ of the first pixel plane from the target are varied with the 
beam momentum so as to accept particles produced in about half a unit of rapidity 
at central rapidity and medium transverse momentum, namely: at 158 $A$\ GeV/$c$\ 
($y_{cm} \simeq 2.9$) $\alpha=40$\ mrad and $d=60$\ cm, at 40 $A$\ GeV/$c$\ 
($y_{cm} \simeq 2.2$) $\alpha=72$~mrad~and~$d=40$~cm.  

\noindent
An array of scintillation counters, placed  10 cm downstream of the target,
provides a fast signal to trigger on the centrality of the collisions. 
The centrality of the Pb-Pb collisions is determined (off-line) by analyzing the 
charged particle multiplicity measured by two stations of silicon strip detectors 
which cover, respectively, the pseudorapidity intervals $2<\eta<3$\ and $3<\eta<4$\ 
for the 158 $A$ GeV/$c$\ set-up, 
and $2<\eta<3$\ and $2.4<\eta<3.7$\ for the 40 $A$ GeV/$c$\ one.
At both energies the triggered fraction of the total nuclear inelastic cross section 
is about 60\%. 

\noindent
In order to estabilish whether in Pb-Pb interactions the multi-strange particle 
production is enhanced with respect to the superposition of  
elementary hadronic interactions,   
reference data are necessary. 
NA57 has collected p-Be data at 40 GeV/$c$\ (whose analysis is ongoing),  
while as reference data at 158 GeV/$c$\ we use both the p-Be and the p-Pb 
sets of data available from the WA97 experiment. 

\section{Data analysis}
The hyperon signals are extracted with a method similar to that used 
in the WA97 experiment~\cite{ReconInWA97}. 
For each particle species we tune the optimal choice of the fiducial  
acceptance window using a Monte Carlo simulation of the apparatus. 
The data are then corrected for geometrical acceptance and for detector and  
reconstruction inefficiencies on a particle-by-particle basis, as  
described in references~\cite{Bruno,Manzari}.   
The stability of the results 
(inverse slopes and extrapolated yields, see later) is checked with 
respect to different choices of the acceptance window.  

\noindent
Finally we determine the double differential cross section 
$\frac{d^2N}{dm_T dy}$\ --- where $y$\ is the {\em rapidity} and 
$m_T=\sqrt{p_t^2+m_{0}^2}$\ is 
the {\em transvers mass} of the particle of rest mass $m_0$\ ---  
and the number of particles per event, i.e. the yield, in the selected 
acceptance window.  
The double differential 
cross section can be parametrized according to the following expression: 
\begin{equation}
\frac{d^2N}{dm_T dy}=A \hspace{1mm} m_T \exp\left(-\frac{m_T}{T_{app}}\right)
\label{eq:mtfit}
\end{equation}
\noindent
where the parameter $T_{app}$, which is referred to as the 
{\em inverse slope} (see section 5 for its physical interpretation), is 
extracted by means of a maximum likelihood fit method.  
By using the parametrization of eq.~\ref{eq:mtfit} we can extrapolate 
the yield measured in the selected acceptance window to a common phase space 
window covering full $p_T$\ and one unit of rapidity centered   
at midrapidity:  
\begin{equation}
Yield=\int_{m}^{\infty} {\rm d}m_{T} \int_{y_{cm}-0.5}^{y_{cm}+0.5} {\rm d}y
  \frac{d^2N}{{\rm d}m_{T} {\rm d}y}.
  \label{eq:yield}
  \end{equation}
\noindent
As a measure of the collision centrality we use the number of wounded nucleons, 
i.e. the nucleons which take part in the initial collisions~\cite{Glauber}. 

\noindent
At 158 $A$\ GeV/$c$\ the multiplicity distribution is divided 
into five centrality classes ($0 ,I,II,III,IV$)~\footnote{The labels I-IV 
were introduced by WA97. The NA57 most peripheral bin is indicated with 
symbol 0 to keep the labelling of WA97 for the other centrality bins.}, 
class $0$\ being the most peripheral and class $IV$\ the most central.  
The average number of wounded nucleons $<N_{wound}>$\ in each class is 
determined 
from the trigger cross section, as described in reference~\cite{Carrer99}. 

\noindent
At 40 $A$\ GeV/$c$, the results on hyperon yields are given 
and compared to those at higher energy for a single centrality class 
corresponding approximately to the most central 42\% of the Pb-Pb interaction 
cross section, which is the equivalent of bins from I to IV at 158 $A$\ GeV/$c$.  

\section{Hyperon yields}
\subsection{Energy dependence}
\begin{figure}[hbt]
\centering
\includegraphics[clip,scale=0.40]{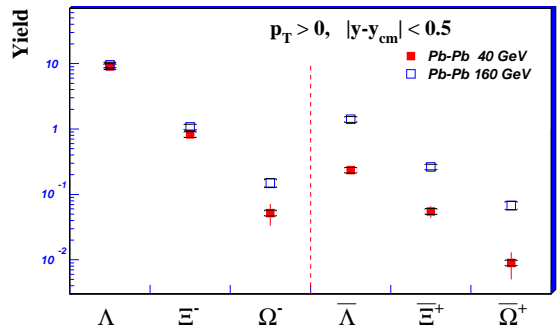}
\caption{Particle yields in 40 and 158 $A$\ GeV/$c$\ Pb-Pb interactions, for 
         about the 42\% most central collisions.  
\label{fig:yie_ener}}
\end{figure}

The yields for the $\La$, $\XI$, $\Omega^-$ hyperons and their 
anti-hyperons are shown in fig.~\ref{fig:yie_ener} at the two beam energies.  
\noindent
Going from 40 to 158 $A$\ GeV/$c$, the $\Lambda$\ and $\XI$\ yields are 
quite similar while the yields of the corresponding anti-particles 
are increased by a factor $\sim5$.  
The $\Omega$\ hyperon production shows a larger increase:   
about a factor 3 for the particle and more than a factor 7 for 
the anti-particle. This picture indicates, as expected, a 
larger baryon density at lower energy.  

\begin{figure}[hctb]
\centering
\includegraphics[clip,scale=0.29]{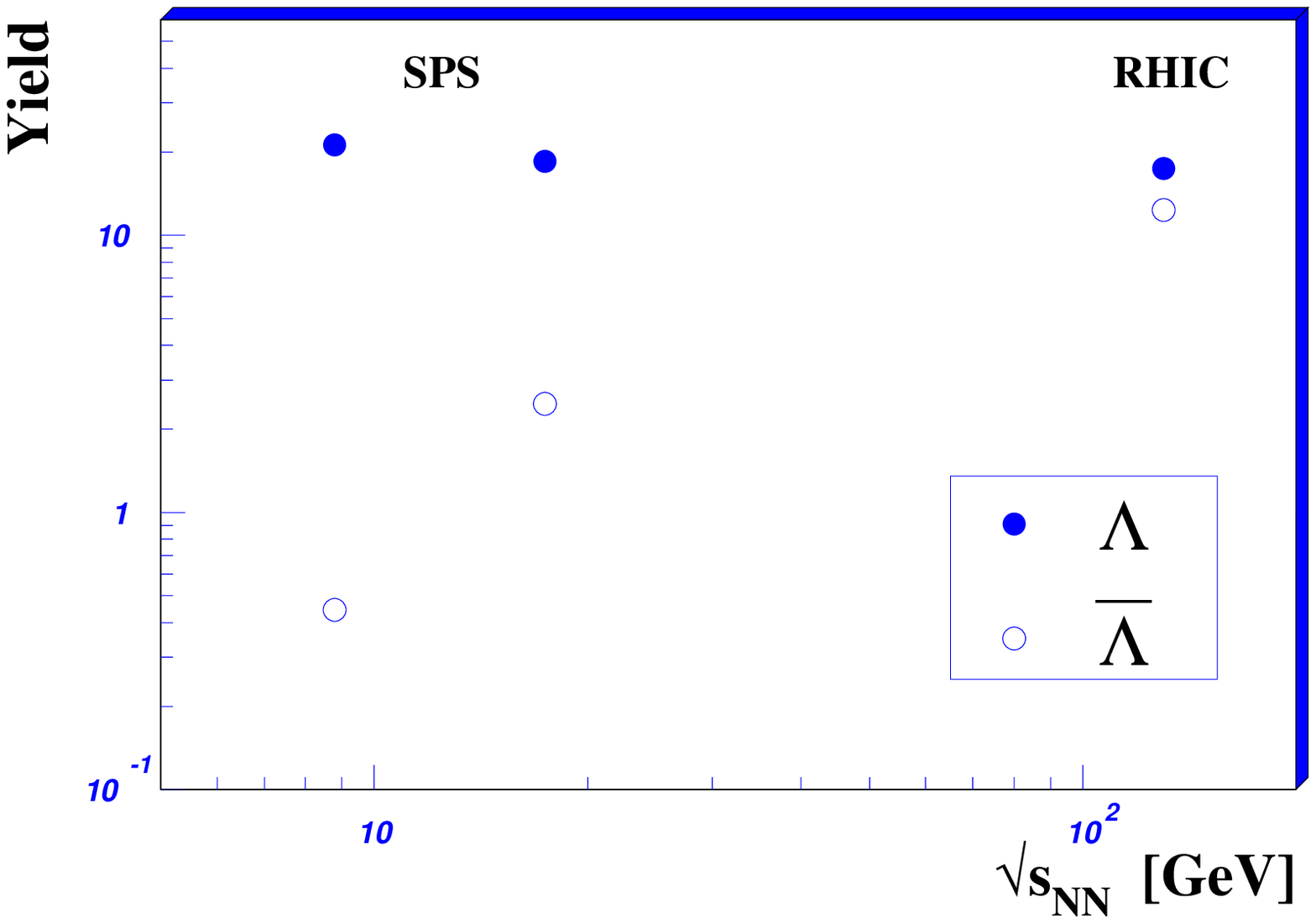}
\includegraphics[clip,scale=0.29]{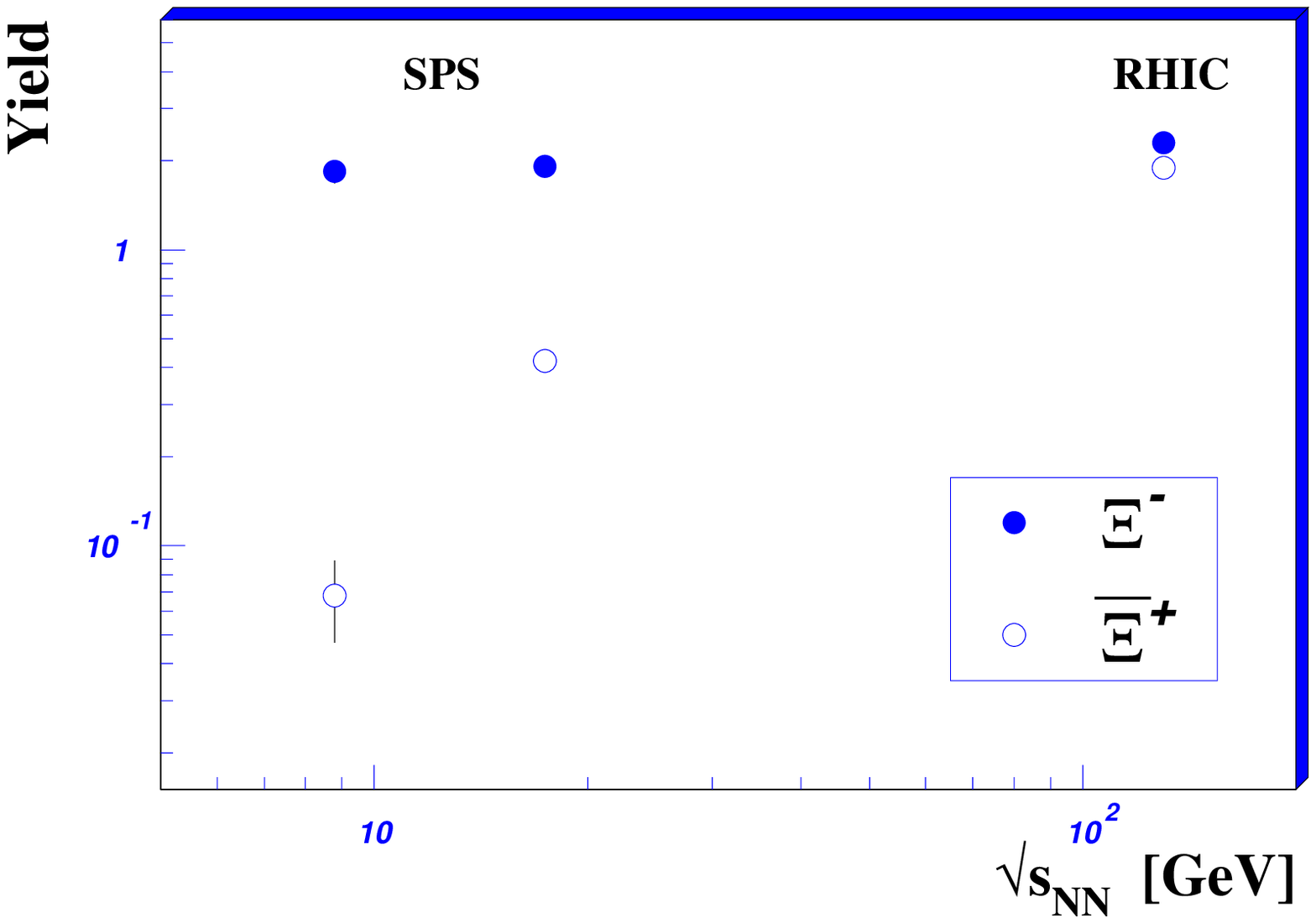}
\includegraphics[clip,scale=0.29]{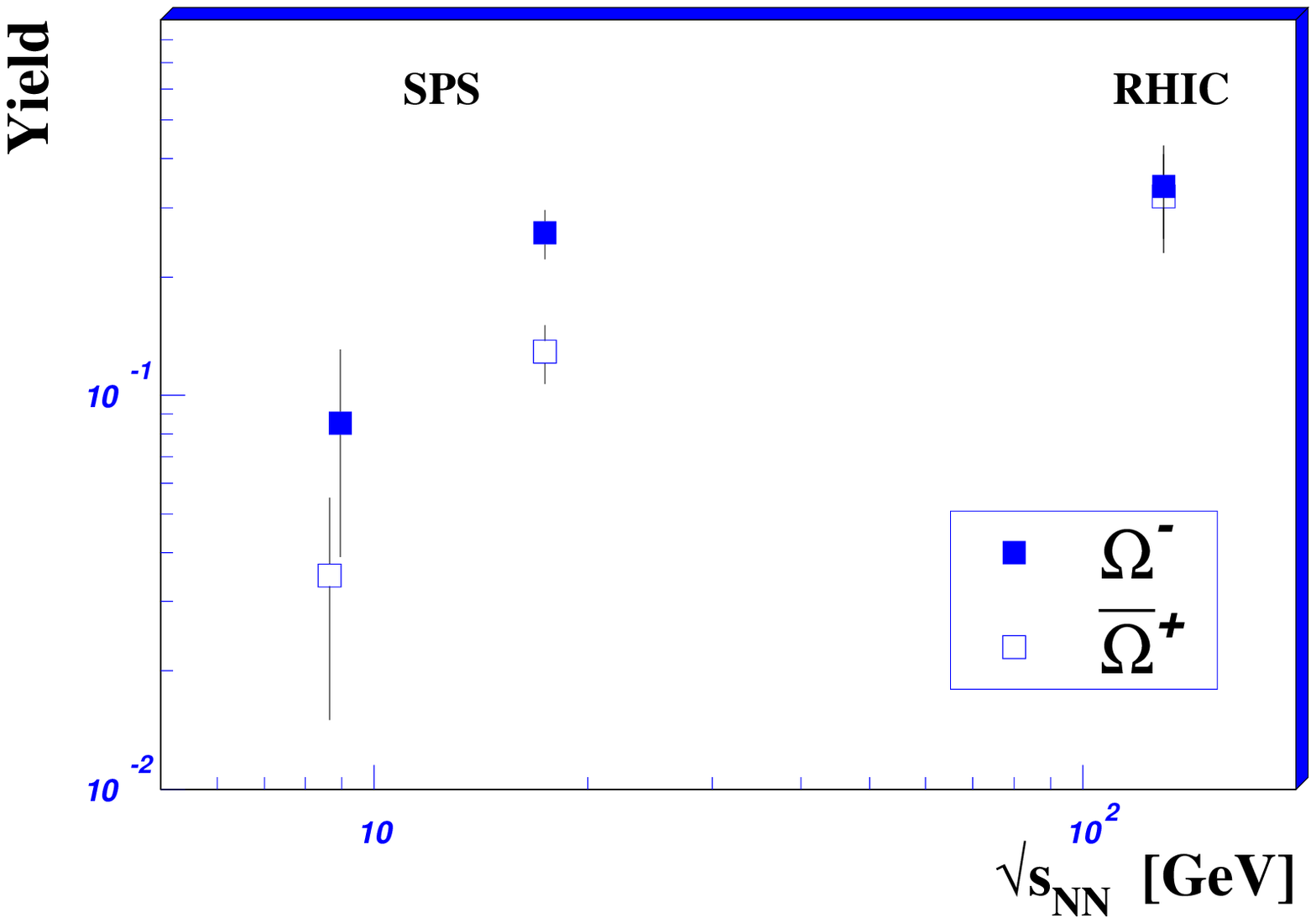}\\
\includegraphics[clip,scale=0.37]{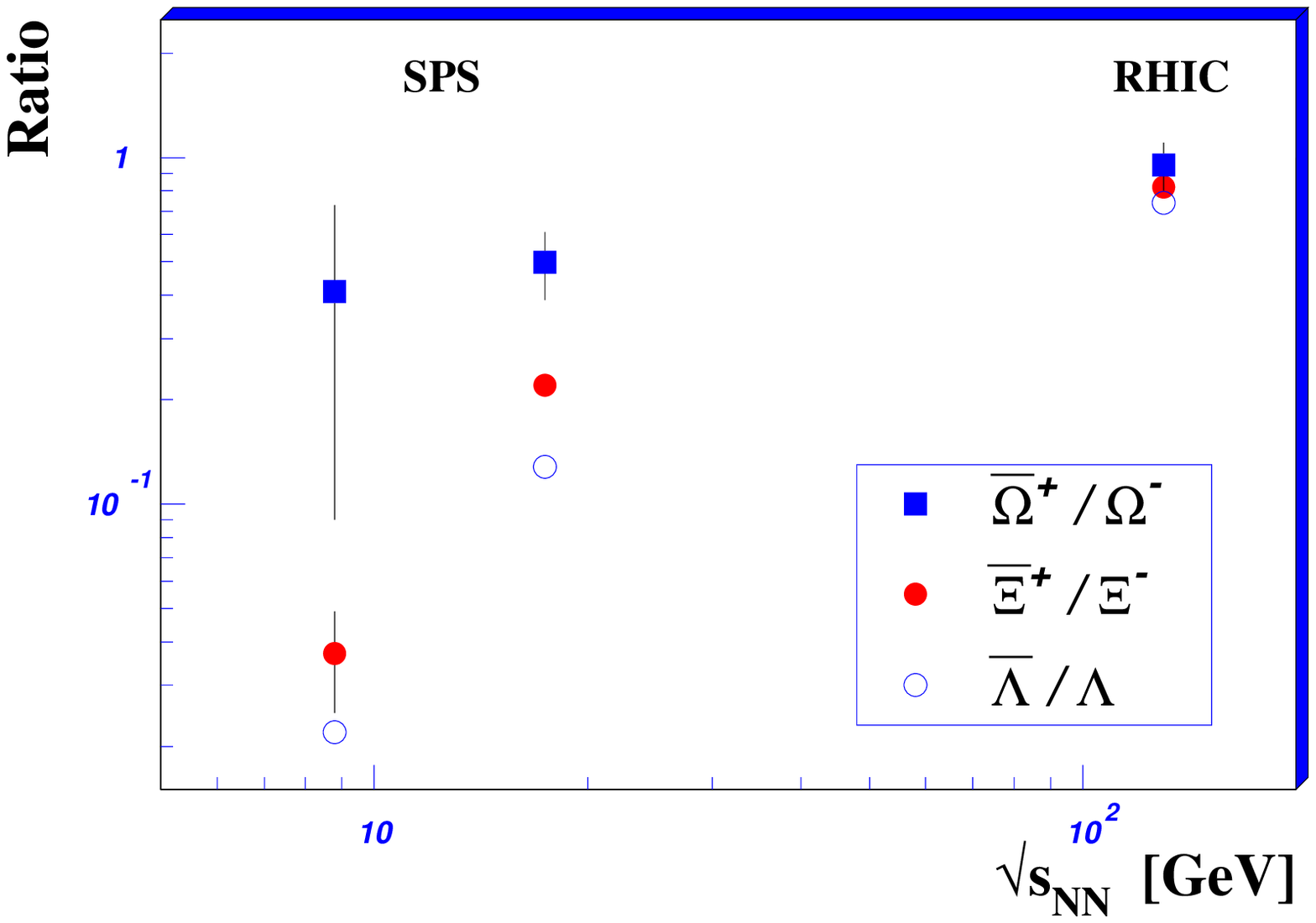}
\caption{Comparison of strange particle yields (top) and anti-hyperon to hyperon 
         ratios (bottom) at SPS and RHIC energies.  
\label{fig:yie_STAR}}
\end{figure}

\noindent
This trend continues when going from SPS to RHIC:  
in fig.~\ref{fig:yie_STAR} the NA57 yields  are compared with those 
obtained in $\sqrt{S_{NN}}=130$\ GeV Au-Au collisions from the STAR 
experiment at RHIC~\cite{STARLa,STARXi,STAROm}  
(NA57 data points at 40 and 160 $A$ GeV/$c$\ correspond   
 to $\sqrt{S_{NN}} = $\ $8.8$\ and $17.3$\ GeV, respectively).   
In order to stay close to the STAR range of 
collision centrality (5\%, 10\% and 14\% most central Au-Au collisions for $\Lambda$, 
$\Xi$\ and $\Omega$, respectively) we restrict here our centrality range to the 
5\% most central Pb-Pb collisions (class IV) in the case of $\Lambda$\ and to the 
12\% most central ones (classes III, IV) for $\Xi$\ and $\Omega$.   
\noindent
The $\La$\ and $\XI$\ yields stay almost constant all along the full energy 
range, whereas an increase with the energy is observed for  
the $\Omega^-$ particle. All the three anti-hyperon yields show a much stronger  
increase with the energy.  

\noindent
From the yields, the ratios between the abundances of 
different particles can be computed. In particular, 
the anti-hyperon/hyperon ratios at SPS and RHIC are larger for higher 
strangeness content of the hyperon (bottom plot of fig.~\ref{fig:yie_STAR}).
All three hyperon ratios also increase as a function of the energy.   
Finally, the energy dependence is found to be weaker for particles  
with higher strangeness content.   
\subsection{Centrality dependence at 158 $A$\ GeV/$c$ }
\begin{figure}[hbt]
\centering
  \includegraphics[clip,scale=0.44]{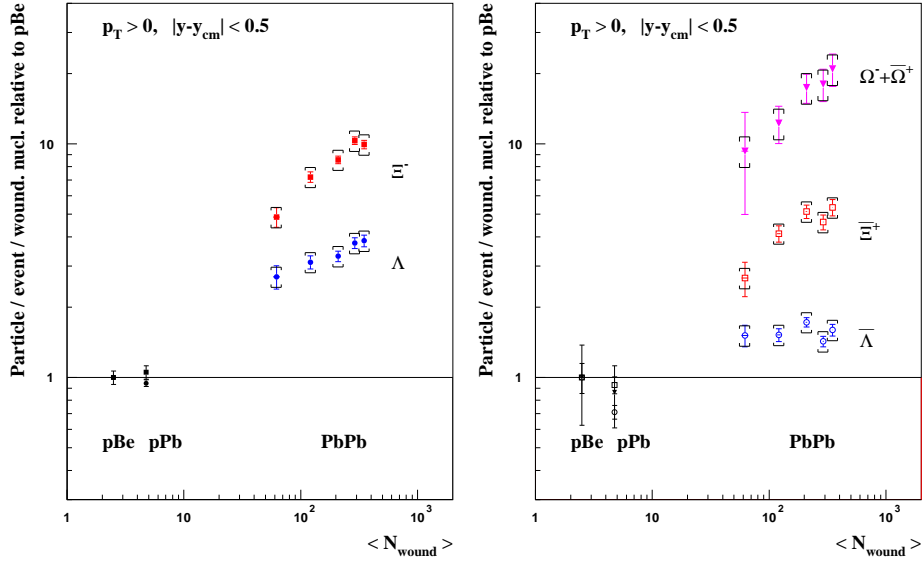}
\caption{Hyperon yields per wounded nucleon per unit of rapidity at central 
         rapidity relative to p-Be as a function of the number of wounded 
	 nucleons. The symbol $_{\sqcup}^{\sqcap}$\  indicates the systematic error. 
\label{fig:partyie}} 
\end{figure}

In fig.\ref{fig:partyie} the yields per wounded nucleon relative to p-Be 
of $\Lambda$, $\Xi$\ and $\Omega$\ and their anti-particles 
in Pb-Pb and p-Pb at 158 $A$ GeV/$c$\  
are plotted as a function of the number of wounded nucleons. 
The  p-Be and p-Pb results are those published by WA97. 
The particles have been divided into two classes, those with 
at least one valence quark in common with the nucleon (left) and those without (right), 
since it is known that the particles of two groups may exhibit different 
production features, e.g. in the rapidity spectra.  
\newline 
The NA57 results confirm the pattern of strangeness enhancements observed by WA97: 
when going from p-Be to Pb-Pb data, 
$\Omega$ are more enhanced than $\Xi$, which are in turn 
more enhanced than $\Lambda$.  
The maximum enhancement is about a factor $20$\ for the 
$\Omega^-+\overline\Omega^+$\ in the most central class.  
\newline
From fig.~\ref{fig:partyie} one can see, within the Pb-Pb sample, 
an increase of the particle yields per wounded nucleon 
with the number of wounded nucleons for all the particles except for the $\Al$. 
\section{Transverse mass spectra at 158 $A$\ GeV/$c$}
\begin{figure}[ht]
\centering
\includegraphics[clip,scale=0.35]{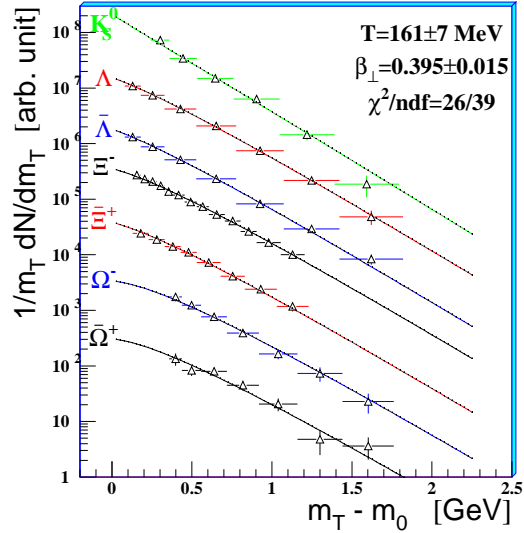}
\caption{Transverse mass spectra of the strange particles measured
         by NA57 with superimposed the result of the best fit with the blast wave
         model.  
\label{fig:spettri}}
\end{figure}

In fig.~\ref{fig:spettri} all the transverse mass spectra of the measured hyperons 
and anti-hyperons and (preliminary) $K_S^0$\ are collected. This sample corresponds 
to 54\% most central Pb-Pb collisions. The values of the 
inverse slope parameters, as calculated according to maximum likelihood 
fits with the function of eq.~\ref{eq:mtfit} (see section 3) are reported 
in table~\ref{tab:InvSlopes}. 
\noindent
\begin{table}[h]
\caption{Inverse slopes parameter ($T_{app}$) of the strange particles
in the full centrality range ({\bf 0-IV}).
\label{tab:InvSlopes}}
\vspace{0.4cm}
\begin{center}
\begin{tabular}{|c||c|c|c|c|c|c|c||}
\hline
{\bf Particle} & {\bf $K_S^0$} & {\bf $\La$} & {\bf $\Al$} & {\bf $\Xi^-$} 
               & {\bf $\overline\Xi^+$} & {\bf $\Omega^-$} 
	       & {\bf $\overline\Omega^+$} \\
\hline
{\bf $T_{app}$\ (MeV)} & {229 $\pm$\ 9} & {289 $\pm$\ 7}  & {287 $\pm$\ 9}
                       & {297 $\pm$\ 5} & {316 $\pm$\ 11} & {280 $\pm$\ 16} 
                       & {324 $\pm$\ 29} \\
\hline
\end{tabular}
\end{center}
\end{table}
\noindent
\newline
Assuming a collective transverse expansion
superimposed to the thermal motion of a fireball locally in thermal equilibrium,
in two extreme  regimes --- namely non-relativistic ($p_T \ll m_0$) and
ultrarelativistic ($m_T \gg m_0$ ) ---
the inverse slope parameter $T_{app}$\ can be simply related to two basic
quantities: the thermal freeze-out temperature ${\bf T}$\ and the
(average) collective transverse flow velocity  ${\bf \beta_{\perp}}$.
In the former regime one should expect an increase of $T_{app}$\ with
the rest mass of the particles: e.g. in reference~\cite{ref1} it has been
derived that $T_{app}={\bf T} + m_0 <{\bf \beta_\perp}^2> $ ;
in the latter regime the inverse slope parameter would be just blue-shifted
with respect to ${\bf T}$\ according to a Doppler formula~\cite{ref2}:
$T_{app}={\bf T}\sqrt{\frac{1+{\bf \beta_{\perp}}}{1-{\bf \beta_{\perp}}}}$.
\newline
For the description of the full $m_T$\ region 
we have employed here the hydro-dynamical model derived in reference~\cite{ref2}.  
In its simplest interpretation, 
when a constant transverse velocity profile is assumed,  
the transverse mass distribution becomes: 
\begin{equation}
\frac{d^2N}{m_Tdm_T dy} \propto m_T 
     K_1\left( \frac{m_T cosh \rho}{\bf T} \right) 
     I_0\left( \frac{p_T sinh \rho}{\bf T} \right)
\label{eq:Blast}
\end{equation}
where $\rho=tanh^{-1} \beta_{\perp}$\ 
and $K_1$\ and $I_0$\ are two modified Bessel functions.  
A simultaneous best fit of eq.~\ref{eq:Blast} to the data points of 
all the measured strange particle spectra successfully describes all 
the distributions with $\chi^2/ndf=26.3/39$, yielding 
the following values 
for the two basic quantities ${\bf T}$\ and ${\bf \beta_\perp}$: 
\[
{\bf T} = 161 \pm 7 MeV \, , \quad \quad {\bf \beta_\perp}=0.395 \pm 0.015  \; .
\nonumber \]
\newline
The particles have been divided again in two groups --- those which share 
quarks in common with the nucleons and those which do not  ---  
and the fit procedure has been repeated separately for the two groups. The results of  
such fits are summarized in table~\ref{tab:Blast1}.
\begin{table}[h]
\caption{Thermal freeze-out temperature $T$\ and average transverse flow velocity 
$\beta_\perp$\ in the full centrality range.   
\label{tab:Blast1}}
\vspace{0.4cm}
\begin{center}
\begin{tabular}{|c||c|c|c||}
\hline
particles & ${\bf T}$ (MeV) & ${\bf \beta_\perp}$ & $\chi^2/ndf$ \\ \hline
{\bf $K_S^0$},  {\bf $\La$},  {\bf $\Xi^-$} & 
$160 \pm 10$ & $ 0.39 \pm 0.02 $ & $ 12.7/15 $ \\ \hline
{\bf $\Al$} , {\bf $\overline\Xi^+$} , {\bf $\Omega^-$} , {\bf $\overline\Omega^+$} &
$167 \pm 16$ & $ 0.39 \pm 0.03 $ & $ 12.2/22 $ \\ 
\hline
\end{tabular}
\end{center}
\end{table}
\noindent
They suggest common freeze-out conditions for the two groups. 
Since the interaction cross-sections for  
the particles of the two groups are quite different, this finding would suggest 
a similar production mechanism 
and limited importance final state interactions (e.g. a sudden thermal freeze-out).   
A similar conclusion concerning the evolution of the system was reached in the 
paper by WA97 studying the HBT correlation functions of negative pions~\cite{HBTpaper}.  
\newline
Finally, we have attempted to study the centrality dependence. We present here very 
preliminary results for three different centralities only, in order to reduce the 
statistical error: our (5\%) most central collision class IV, 
classes 0 and $I$\ merged (most peripheral group) and 
classes II and III merged (intermediate centrality). In table~\ref{tab:Blast3} are 
reported the fit parameters and the centrality ranges in terms of residual percentage of 
the total inelastic cross-section. The suggested trend is as follows:  
the more central the collisions the larger the transverse collective flow and the lower 
the final thermal freeze-out temperature. 
\begin{table}[h]
\caption{Thermal freeze-out temperature $T$\ and average transverse flow velocity
$\beta_\perp$\ as a function of centrality.  
\label{tab:Blast3}}
\vspace{0.4cm}
\begin{center}
\begin{tabular}{|l||c|c|c||}
\hline
Centrality & ${\bf T}$ (MeV) & ${\bf \beta_\perp}$ & $\chi^2/ndf$ \\ \hline
0-I \quad $23 \div54 \%$    & $163 \pm 16$ & $ 0.35 \pm 0.03 $ & $ 37.8/37 $ \\ \hline
II-III $5 \div 23 \%$ & $167 \pm 13$ & $ 0.40 \pm 0.03 $ & $ 32.6/37 $ \\ \hline
IV  \quad $ 0 \div5 \%$     & $131 \pm 10$ & $ 0.47 \pm 0.02 $ & $ 37.4/37 $ \\
\hline
\end{tabular}
\end{center}
\end{table}
\noindent
The higher temperature for peripheral collisions 
could be due to the shorter time duration 
of the expansion:   
the sytem would have less time for cooling before of the final decoupling. 
The results for our most central class are consistent within the errors with those   
in the same centrality range   
from a similar analysis by NA49~\cite{BlastNA49}.  
\section{Conclusions and outlook}
Results from NA57 on $\La$, $\Xi$\ and $\Omega$\ production 
in Pb-Pb collisions at both 40 and 158 $A$ GeV/$c$ have been reported.   
The energy dependence of the hyperon and anti-hyperon yields  is compatible with   
a decrease of the baryon density with increasing energy.   
Going from low (40 $A$~GeV/$c$) to top (158 $A$~GeV/$c$) SPS beam momentum  
the $\Lambda$\ and $\XI$\ yields stay roughly constant while 
the production of the $\Omega^-$\ and all the anti-hyperons 
increases significantly with the collision energy; 
this trend continues up to the RHIC energy.  
At 158 $A$\ GeV/$c$\ the centrality dependence of  
the yields per wounded nucleon confirms the pattern of  
enhancements relative to p-Be observed by WA97: the enhancement increases with the  
strangeness content of the particle, up to a factor $\approx 20$\ for 
$\Omega^- + \overline\Omega^+ $. All the particles except the $\Al$\ show an 
increase of the enhancements, when going from peripheral   
to central Pb-Pb collisions: a saturation may be possible for the 
two -- three most central bins.   
The ongoing data analysis on p-Be collisions will soon provide results on 
the pattern of strange particle enhancements at 40 $A$\ GeV/$c$. 
\newline
The preliminary analysis of the transverse mass spectra at 158 $A$\ GeV/$c$\ 
in the framework of a hydro-dynamical model with a constant velocity profile 
suggests  that after the collision the system expands explosively, with a  
tranverse velocity of about one half of the speed of light;     
the system then freezes-out when the temperature is of the order of 150 MeV. 
Finally the preliminary results on the centrality dependence of the 
expansion dynamics indicate that with increasing centrality 
the transverse flow velocity increases and the final temperature decreases.

\end{document}